# Periodic elastic nanodomains in ultrathin tetrogonal-like BiFeO$_3$ films

Zhenlin Luo,[1,*,†] Zuhuang Chen,[2,§,†] Yuanjun Yang,[1] Heng-Jui Liu,[3] Chuanwei Huang,[2] Haoliang Huang,[1] Haibo Wang,[1] Meng-Meng Yang,[1] Chuansheng Hu,[1] Guoqiang Pan,[4] Wen Wen,[5] Xiaolong Li,[5] Qing He,[5] Thirumany Sritharan,[2] Ying-Hao Chu,[3] Lang Chen,[2,6,‡] and Chen Gao[1,#]

[1]*National Synchrotron Radiation Laboratory and CAS Key laboratory of Materials for Energy Conversion, Department of Materials Science and Engineering, University of Science and Technology of China, Hefei 230026, China*

[2]*School of Materials Science and Engineering, Nanyang Technological University, Singapore 639798, Singapore*

[3]*Department of Materials Science and Engineering, National Chiao Tung University, Hsinchu 30010, Taiwan*

[4]*National Synchrotron Radiation Laboratory, University of Science and Technology of China, Hefei 230026, China*

[5]*Shanghai Synchrotron Radiation Facility, Shanghai Institute of Applied Physics, Chinese Academy of Sciences, Shanghai 201204, China*

[6]*Department of Physics, South University of Science and Technology of China, Shenzhen, China, 518055, P.R. China*

† Z. L. Luo and Z. H. Chen contributed equally to this work.

**\*zlluo@ustc.edu.cn, §glory8508@gmail.com, ‡chenlang@sustc.edu.cn,**

**#cgao@ustc.edu.cn**

**ABSTRACT**

We present a synchrotron grazing incidence x-ray diffraction analysis of the domain structure and polar symmetry of highly strained $BiFeO_3$ thin films grown on $LaAlO_3$ substrate. We revealed the existence of periodic elastic nanodomains in the pure tetragonal-like BFO ultrathin films down to a thickness of 6 nm. A unique shear strain accommodation mechanism is disclosed. We further demonstrated that the periodicity of the nanodomains increases with film thickness but deviates from the classical square root law in ultrathin thickness regime (6 - 30 nm). Temperature-dependent experiments further reveal the disappearance of periodic modulation above ~ 90 °C due to a $M_C$-$M_A$ structural phase transition.

Keywords: $BiFeO_3$, Ultrathin multiferroic films, Periodic nanodomains, Grazing incidence X-ray diffraction, Epitaxial strain

## I. INTRODUCTION

With the miniaturization trend of functional devices, nanoscale epitaxial ferroic (ferroelectric, ferromagnetic, ferroelastic, *et al.*) thin films have gained great attention in recent years due to their fundamental physics and practical applications in such as sensors and information storage.[1] Domains form in ferroic films to minimize the free energy of depolarizing (or demagnetizing) fields and elastic film-substrate interaction (epitaxial strain).[2] For instance, epitaxial strain due to lattice mismatch between film and underlying substrate creates a driving force for the formation of regular elastic (non-180°) domains in ferroic films grown on single crystal substrates.[3-7] Such elastic domain was predicted theoretically by Roitburd in 1976,[3] and was experimentally observed in tetragonal $PbTiO_3$,[8] $Pb(Zr_xTi_{1-x})O_3$,[9] $BaTiO_3$,[10] rhombohedral $BiFeO_3$ (BFO),[11] orthorhombic $NaNbO_3$,[12] and Aurivillius layered $SrBi_2Ta_2O_9$,[13] epitaxial ferroelectric films. Domain structure plays a crucial role in determining the physical properties of ferroic thin films.[2] Furthermore, recent studies have revealed that domain walls themselves present novel functionalities,[14] such as enhanced conductivity,[15] and magnetism.[16] Therefore, understanding and controlling the ferroic domain structure, especially in ultrathin films, are of importance to realize the practical applications. However, taking ferroelectrics for example, most previous experimental studies focused on relatively thick films with dense domain structure, where domain sizes are much smaller than the film thickness;[7] and the domain structure and domain size evolution with film thickness in ultrathin films have been less thoroughly investigated. This is mainly because probing the domain structure in ultrathin films is challenging. Scanning probe techniques do not readily provide sufficient resolution to detect domain structure in ultrathin ferroelectric films because of the weak piezoelectric response and fine domain feature of the ultrathin films. Transmission electron microscopy has been used instead, but it is destructive and might change the electric and mechanical boundary conditions and thus may not provide the original structural information, especially for the highly-strained films.[17] Recently, synchrotron grazing incidence x-ray diffraction (GIXRD) techniques have been used successfully to probe nanoscale 180° ferroelectric stripe domains in PbTiO$_3$ ultrathin films as the periodic nanodomains produce satellite peaks in x-ray scattering.[18-20] The distribution and orientation of these satellites from nanodomains in reciprocal space can provide rich information on the domain structure and polar symmetry in ferroelectric ultrathin films.[18-21]

Among ferroic materials, BFO is of particular interest because of its lead-free nature, large ferroelectric polarization, robust piezoelectricity, room temperature multiferroic properties, and relatively small band gap.[22] The crystal and domain structures of rombohedral BFO films have been extensively studied in the past decade.[22, 23] Recently, a novel 'super *tetragonal*' BFO phase with a giant axial ratio and a huge spontaneous polarization was predicted by theoretical studies,[24] and experimentally confirmed in highly strained BFO films grown on LaAlO$_3$ (LAO),[25-27] LaSrAlO$_4$,[28] NdAlO$_3$,[29] and YAlO$_3$ substrates.[26] The crystal structure of this 'super tetragonal' BFO at room temperature was determined to be tetragonal (T) -like $M_C$-type monoclinic with *Pm* or *Pc* symmetry.[27, 30, 31] Besides the giant polarization,[32] an near-room-temperature multiferroic phase transition from the $M_C$ phase to a high-temperature T-like monoclinic $M_A$ phase, was reported for T-like BFO films more recently, implying potential applications of this novel T-like phase.[33-36] However, despite recent intensive studies on T-like BFO films,[25-39] the detailed domain structure and its temperature- and thickness-dependence, and the strain-accommodation mechanism in these highly-strained films, especially in ultrathin pure T-like films, have not yet been fully understood. This hinders full understanding of the structure-property relationship and the accuracy of inferences on the magnetoelectric coupling in the newly discovered 'super *tetragonal*' phase.

In this paper, we report observations of unique periodic elastic stripe nanodomains and their thickness- and temperature-dependence using synchrotron x-ray diffraction in a series of sub-30-nm-thick T-like BFO ultrathin films grown on (001)-oriented LAO substrates. We also demonstrate that the stripe domain period increases with the film thickness but deviates from the classical square root law, and the $M_C$ phase transforms to a pure T-like $M_A$ phase at ~90 °C simultaneously with the disappearance of the periodic domains.

## II. EXPERIMENTAL METHODS

Epitaxial BFO thin films with thicknesses ranging from 2 nm to 30 nm were grown on (001)-oriented LaAlO$_3$ (LAO) single crystal substrates using pulsed laser deposition at 700 °C under an oxygen pressure of 100 mTorr.[28] The crystallographic structure of the films was studied using high resolution synchrotron x-ray diffraction from two sources: BL14B1 beamline of the Shanghai Synchrotron Radiation Facility (SSRF, λ = 1.2398 Å) for conventional reciprocal space mappings (RSMs) and U7B beamline of the Hefei National Synchrotron Radiation Laboratory (NSRL, λ = 1.537Å) for GIXRD studies.[40] The RSMs were plotted in reciprocal lattice units (r.l.u.) of the LAO substrate (1 r.l.u. = 2π/3.789 Å$^{-1}$) and the diffraction intensity was indicated by different colors (low to high: blue-green-yellow-red-grey). Film thickness was determined by analyzing synchrotron x-ray reflectivity data.[40]

## III. RESULTS AND DISCUSSION

Previous studies have revealed that pure T-like phase only occurs in BFO films of thickness less than 30 nm grown on LAO, while a larger thickness leads to multiphase coexistence due to strain relaxation.[26, 37] Here, all films display atomically flat terraces with single-unit-cell steps in the AFM topography.[37, 40] **Figure 1** shows typical RSMs in pseudo-cubic coordinates around (002) and ($0\bar{1}3$) reflections for a 10-nm-thick film. Only peaks from the substrate and the BFO film with out-of-plane lattice parameter of 4.64 Å were detected, indicating that the film is pure tetragonal-like. The presence of thickness fringes along the $L$ direction indicates that the film is very smooth, which is consistent with the topography data.[37, 40] As shown in Fig. 1(b), satellite peaks with identical spacing are observed around the $0\bar{1}3$ Bragg peak of the film, implying the existence of periodic domains in the film. Further, both the central and the satellite peaks show thickness fringes, indicating that the periodic domains extend through the entire film. The lack of satellite peaks around pure out-of-plane reciprocal lattice points ($H = K = 0, L \neq 0$), such as (002) shown in Fig. 1(a), suggests that the relative positional shift of atoms in adjacent domains lies horizontally without any out-of-plane component.

In order to get more information about the in-plane domain structures, GIXRD technique was adopted to map out the in-plane reciprocal space. **Figure 2** shows typical RSMs around various in-plane reciprocal lattice points for the 10 nm film, measured in grazing incidence geometry. These RSMs are proportionally enlarged to show details. Obviously, satellites are observed around all ($HK$0) reciprocal lattice points. The satellites around 010 / 020 Bragg peaks are aligned along both in-plane diagonal directions ([110] and [$\bar{1}$10]), while the satellites around 220 / $\bar{2}$20 peaks lie only along one direction ([$\bar{1}$10] or [110]). Analysis of the diffraction data reveals that the reflections originate from two sources: domain variants of T-like BFO and periodic alignment of these domains.[41] For instance, comparing the ($\bar{1}$10) RSM with the ($\bar{2}$20) one, it could be deduced that: 1) the outer major diffraction spots, fallen on lines converge to the origin, are reflections from the same family of crystallographic planes but with different orders, and 2) the inner satellite spots with identical spacing stand for a superlattice-like structure of periodic nanodomains. The oval profile of domain reflections indicates that stripe domains aligned along [$\bar{1}$10] direction and the splitting can be attributed to the twinned structure of the LAO substrate. The satellites spacing value yields an in-plane periodicity of ~30 nm. After detailed analysis of all RSMs, the origins of all reflection spots are identified and marked in Fig. 2. Since the diffraction intensity profile shown in these RSMs comes from the convolution integral of the diffractions from crystal domains as well as lattice modulation due to the periodic domains, peak overlap is inevitable, which is especially obvious in the (010) and (020) RSMs.

Fig. 2 shows that the modulations due to periodic nanodomains are along <110> directions, which are perpendicular to the domain wall orientations revealed by PFM.[27] This can be interpreted by a model of periodic $M_C$ domains with antiphase in-plane rotations, as shown in **Figure 3**. Figs. 3(a-c) show the orientation relationship between the in-plane lattices of the film and the substrate. For quantitative analysis, in-plane lattice parameters of $a$ = 3.833(3) Å, $b$ = 3.747(3) Å extracted from a 24-nm-thick BFO film are used, since the lattice parameters of the pure T-like films keeps almost constant.[28,40] The in-plane rotations of adjacent $M_C$ domains are determined to be ±0.661° (±0.005°) from the relative shift of BFO (020) reflection from the LAO [010] axe.[40] Calculation shows that the film lattice is coherent with that of the LAO substrate along one diagonal direction [110]/[$\bar{1}$10] after rotation, where the lattice constants match by ~99.9% and the angle difference is only 0.01°. The domain pattern observed in the reciprocal space, as illustrated in Fig. 3(d), is in good agreement with the fourfold in-plane alignment symmetry of the $M_C$ domain variants and two choices of in-plane rotation proposed here. The domain structure is composed of a rotated domain (marked as domain A in Fig. 3) and another adjacent anti-phase equal-angle rotated domain (marked as domain B in Fig. 3). The observed in-plane periodic modulation along [110]/[$\bar{1}$10] could be attributed to one-dimensional periodic alignment of such $M_C$ domain twins with domain walls lying along [$\bar{1}$10]/[110]. Two cases ("Head to Tail" and "Head to Head") of possible periodic domain structures with domain walls along <110> directions in real space and the corresponding simulated satellite around ($\bar{1}$10) in reciprocal space are illustrated in Fig. 3(e) for the $M_C$ phase, where each case assumes the presence of four domain structures due to the four-fold symmetry of (001)-oriented pseudo-cubic substrate. The observed satellites in ($\bar{1}$10) RSM for our T-like BFO films are consistent with the "head to tail" case where domain walls are uncharged. The satellites due to periodic domains are only evident around reciprocal lattice points with in-plane component (H ≠ 0 or K ≠ 0). This suggests that the periodic nanodomains are pure elastic ones and have the same out-of-plane polarization component, which is consistent with the out-of-plane PFM image that shows a uniform contrast.[27, 37, 40] The periodic elastic nanodomain pattern observed here is different from the ferroelectric 180° stripe domain structure found in PbTiO$_3$ ultrathin films in order to reduce the depolarization energy.[18-20] In addition, it is important to note that the satellite spots are not caused by the twinned substrate, because twin size of the LAO substrate is much larger, in the order of tens of micrometers, than the modulation period exhibited in the RSMs. This claim is further confirmed by supplementary experiments that we found no existence of lattice modulation from bare LAO substrate after similar heat treatment.[40]

Based on the above discussion, it is clear that the GIXRD RSMs obtained here present a sum effect of the $M_C$ domain variants and the periodic modulation of nanodomain twins, which discloses a unique strain accommodation mechanism of the highly strained pure T-like BFO film on LAO, i. e., stripe $M_C$ BFO domains with small antiphase in-plane rotations lying periodically in the film along one <110> direction with uncharged domain walls. The shear strain energy due to the unit-cell rotation could be further diminished by a hierarchical structure constructed from the alternation of the two configurations of periodic $M_C$ domains, as shown in **Figs. 3(a) and 3(b)**. A similar hierarchical structure of four twins has been predicted by Roytburd in 1993 for tetragonal ferroelectric thin films.[4, 5]

To investigate the thickness-dependent domain structure of T-like BFO films, GIXRD was performed on films with various thicknesses. Overall the profiles of the in-plane RSMs were similar to that of the 10-nm BFO film. Typical RSMs of the 24-nm and 30-nm films are shown in **Fig. 4**. It could be noted that these BFO films are pure T-like except for the 30 nm film, where the appearance of a weak reflection spot at the bottom of Fig. 4c signifies a minority rombohedral-like phase. It also should be noted that, with the increasing film thickness, the relative intensities of domain reflections become larger while those of the periodic modulation reflections become smaller with a smaller interval. This can be attributed to increasing domain size leading to bigger modulation periodicity in thicker films. Moreover, as clearly shown in the ($\bar{1}10$) RSMs, positional shift of the domain reflections is not obvious (<<1%), indicating unchanging lattice constant of the $M_C$ phase. The stripe domain period $D$ in the pure T-like BFO films are plotted in Fig. 4e in a log-log scale as a function of film thickness $d$. Notably, for the T-like films with thickness below 30 nm, the domain period $D$ scales with the film thickness $d$, obeying a power law $D \propto d^{\gamma}$, with a scaling exponent $\gamma \approx$ 0.75±0.05. In epitaxial ferroic films, the stripe domain periodicity generally experiences a non-monotonous thickness-dependent evolution,[4, 14, 42-44] including three distinct regions: (I) a classical square root law ($\gamma = 0.5$) for thick films where the film thickness $d$ is much larger than the domain periodicity $D$, (II) a deviation from the square root law for intermediate film thickness, and (III) an increase exponentially in ultrathin films when decreasing film-thickness. For thick ferroic films, where the domain size is much smaller than the film thickness (i.e., dense domain structure), the domain width scales with the square root of film thickness, which has been clarified in ferromagnetics by Landau, Lifshitz and Kittel,[45] ferroelectrics by Mitsui and Furuichi,[46] and ferroelastics by Roitburd.[3] For intermediate films (region II), the domain size becomes comparable to the film thickness, which leads to a significant increase of the electrostatic interaction from the films surfaces. Consequently, the domain width scaling behavior starts to diverge from the classical law.[43] This is in reasonable agreement with our observed deviation of square root scaling behavior in the pure T-like BFO films below 30 nm, where the size of the $M_C$ domain is comparable with the film thickness. Furthermore, no satellites due to periodic domains were detected in a 2 nm BFO ultrathin film, indicating that the film is likely to be monodomain (region III).

Thermally driven evolution of the T-like phase and its domain pattern was investigated by temperature-dependent GIXRD. The results obtained in a 12-nm-thick BFO film are shown in **Figure 5**. Figs. 5(a-f) present the ($\bar{1}10$) and (010) RSMs at 100 ℃, 80 ℃ and 25 ℃, respectively. The outer domain reflection peaks disappear at 100 ℃ simultaneously with the super-lattice satellites, while the central one becomes stronger in intensity, indicating a structural phase transition occurring between 80 ℃ and 100 ℃. Detailed rocking curves of ($\bar{1}10$) reflection (2θ=33.4°) were measured at different temperatures and shown in Fig. 5(g). Below 85 ℃, the dependence of satellite spacing on temperature is relatively weak. At ~85 ℃, there are abrupt changes in both intensity and position of the satellite peaks and outer reflection peaks of the $M_C$ phase. Eventually all these peaks vanish at 90 ℃. Some supporting evidence is available in recent reports,[33-36, 47, 48] which show that the film becomes T-like $M_A$ phase above 90 ℃. Such $M_C$ - $M_A$ structural transition near room temperature has been detected in very recent studies by different techniques, such as Raman spectroscopy,[33] neutron diffraction,[34] TEM,[48] and PFM.[36, 47] Our experimental result presented here unambiguously shows that both the crystal and domain structures of T-like BFO undergo a phase transition at ~90 ℃. Since the crystal symmetry and domain structure (including domain walls) play an important role on the unique physical properties of BFO, this near-room-temperature structural and domain pattern's change will lead to various controllable functionalities in BFO, such as magnetoelectric properties. The disappearance of satellite spots in the XRD pattern above the $M_C$ - $M_A$ phase transition temperature further confirms that the satellite peaks in the GIXRD patterns originates from periodic polar domains rather than other periodic defects, such as dislocations.

## IV. CONCLUSION

In summary, high resolution synchrotron XRD studies reveal the existence of periodic elastic nanodomains in pure T-like BFO/LAO epitaxial films with thickness down to 6 nm. A unique strain accommodation mechanism is disclosed, where the adjacent T-like BFO domains adopt small antiphase in-plane rotations to match better with each other as well as with the substrate along the diagonal. The domain size increases with film thickness at a power law constant ~0.75 and deviates from the classical square root law. Temperature-dependent studies show that the periodic nano-domain modulation vanishes when a structural transition occurs from T-like $M_C$ phase to T-like $M_A$ phase at ~90 ℃. This correlates the evolution of domain structure with intriguing near room-temperature magnetic and ferroelectric property changes discovered very recently in highly-strained BFO films.[21-24] These findings not only enrich the understanding of the nature of T-like BFO, but also underscore the unique power of synchrotron XRD for the determination of polar symmetry, domain structure and strain-accommodation mechanism in ferroic ultrathin films.

**Acknowledgements**

We are grateful to Dr. Ping Yang and Dr. Dillon Fong for useful discussions. This work was supported by the National Basic Research Program of China (973 program) (2010CB934501, 2012CB922004), the Natural Science Foundation of China (11004178, 11179008, 51021091, 50832005), and the Chinese Universities Scientific Fund. ZC acknowledges the support from Ian Ferguson Postgraduate Fellowship. LC acknowledges the supports from the Singapore National Research Foundation under CREATE program: Nanomaterials for Energy and Water Management, 9th Thousand Talent Program of China and a start-up fund from SUSTC. The work at NCTU is supported by the National Science Council, R.O.C (NSC-101–2119-M-009–003-MY2), Ministry of Education (MOE-ATU 101W961), and Center for interdisciplinary science of National Chiao Tung University. The authors thank beamline U7B of NSRL and beamline BL14B1, BL08U of SSRF for providing the beam time.

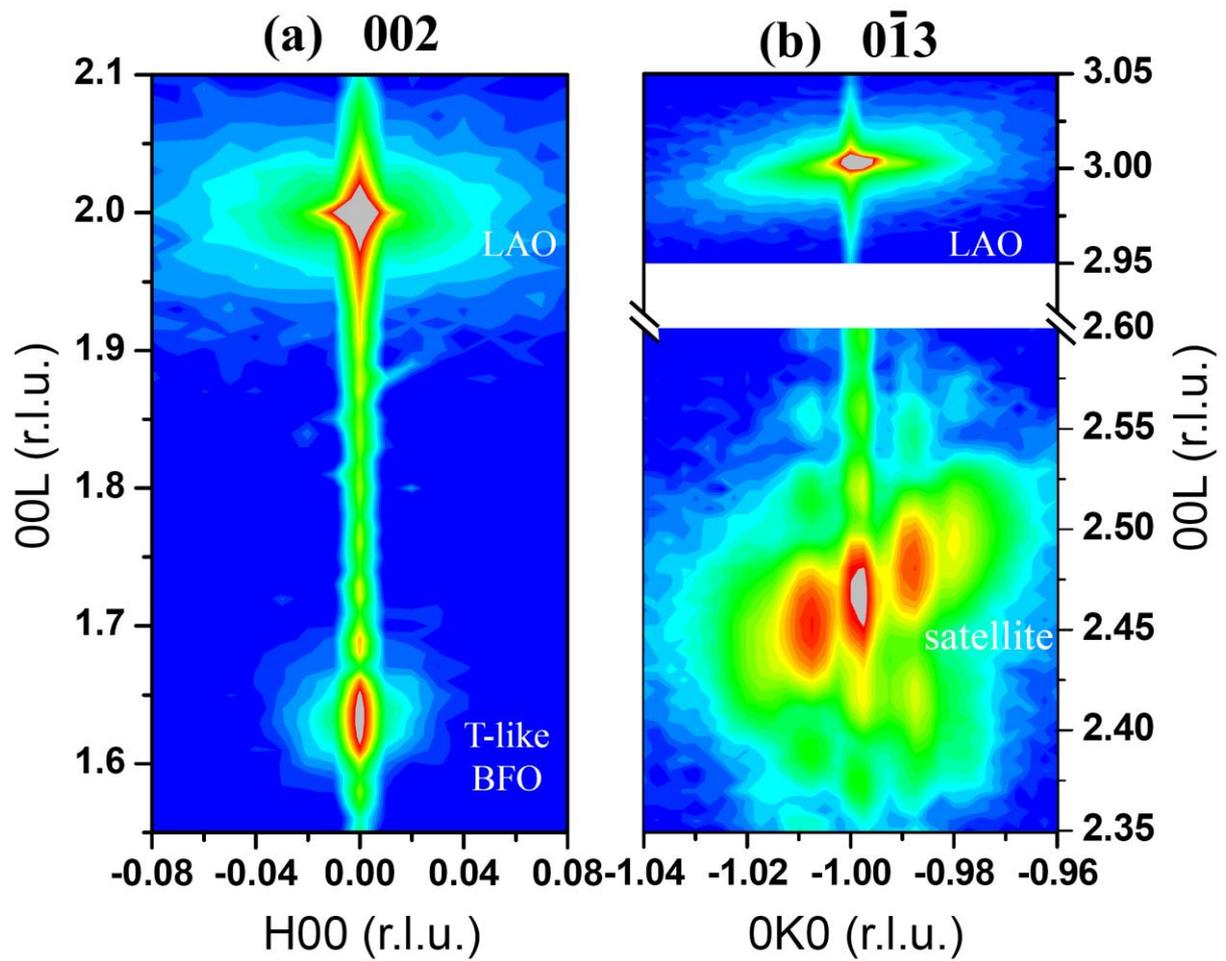

**Figure 1.** (color online) RSMs around (002) and ($0\bar{1}3$) reflections for a 10-nm-thick BFO film on LAO. (Intensity from low to high: blue-green-yellow-red-grey).

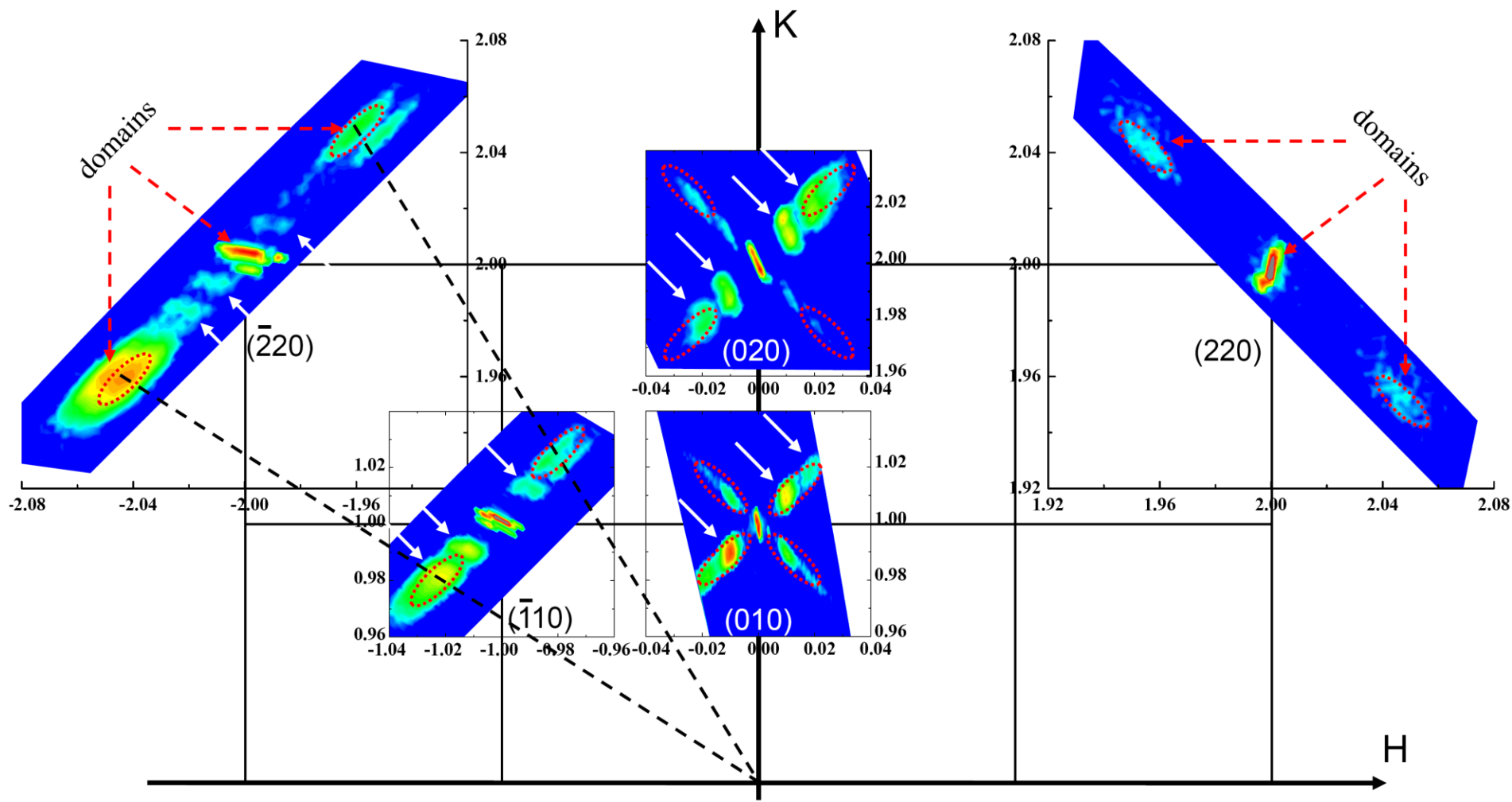

**Figure 2.** (color online) In-plane RSMs of the 10-nm-thick BFO film. These RSMs are proportionally enlarged. Red arrows and red circles indicate the reflections of domain variants while white arrows indicate those of periodic modulation. (Intensity from low to high: blue-green-yellow-red-grey).

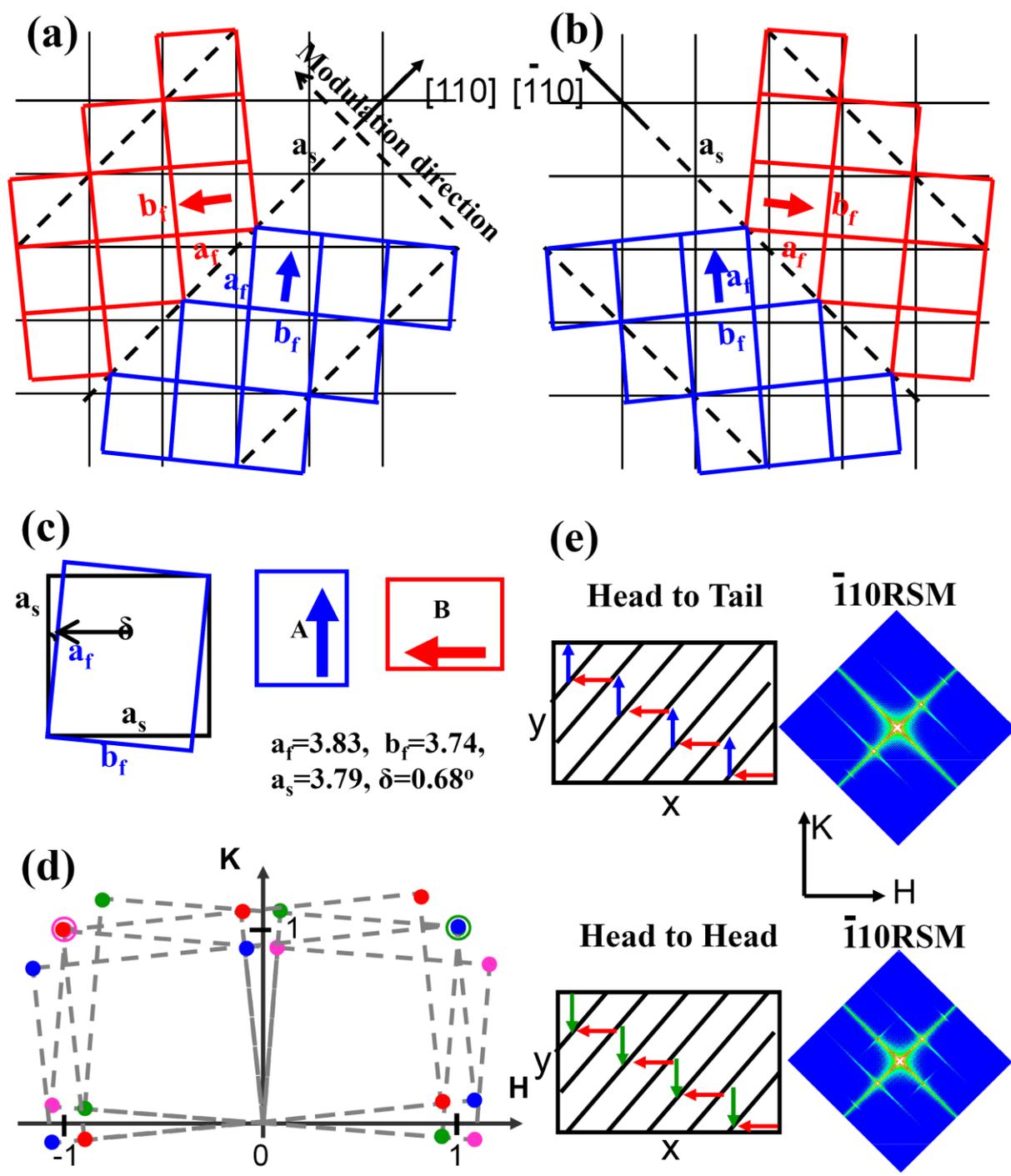

**Figure 3.** (color online) (a-c) Schematic diagraph of lattice matching between $M_C$ domain twins and the LAO substrate. Blue: $M_C$ domain A; Red: $M_C$ domain B; Black: substrate lattice. Blue and red Arrows represent the in-plane polarization vectors. (d) Sketch of all $M_C$ domain variants in reciprocal space (L=0 plane). (e) Illustration of two cases of possible periodic domain structures of ferroelectric $M_C$ phase in real space and the corresponding simulated satellites in ($\bar{1}10$) RSMs, on (001)-oriented pseudo-cubic substrate.

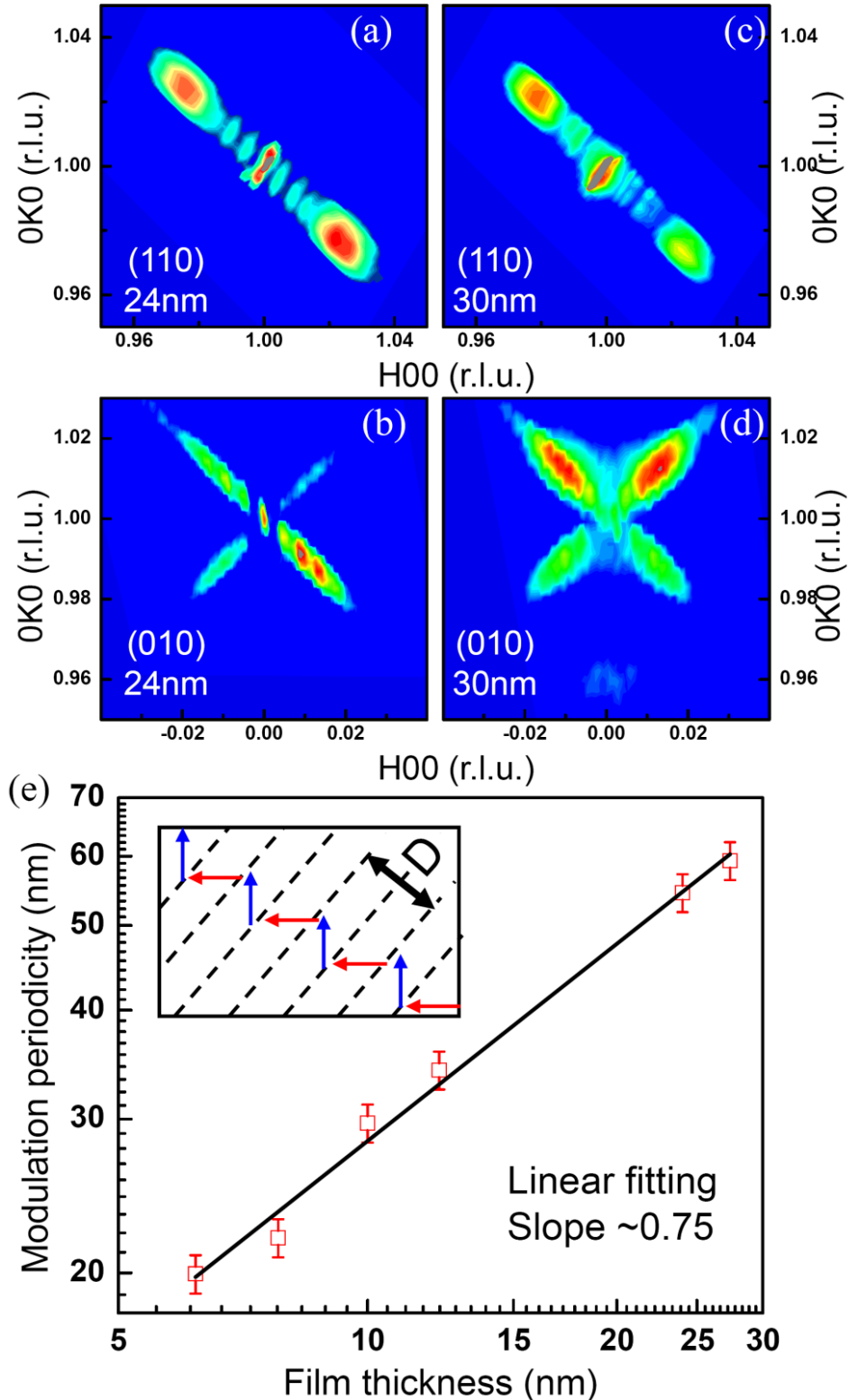

**Figure 4.** (color online) In-plane RSMs around (010) and (110) reflections for the 24-nm (a,b) and 30-nm (c,d) BFO films. (Intensity from low to high: blue-green-yellow-red-grey). (e) Periodicity of in-plane stripe domain as a function of film thickness for the T-like BFO films at room temperature. The domain pattern is illustrated in inset.

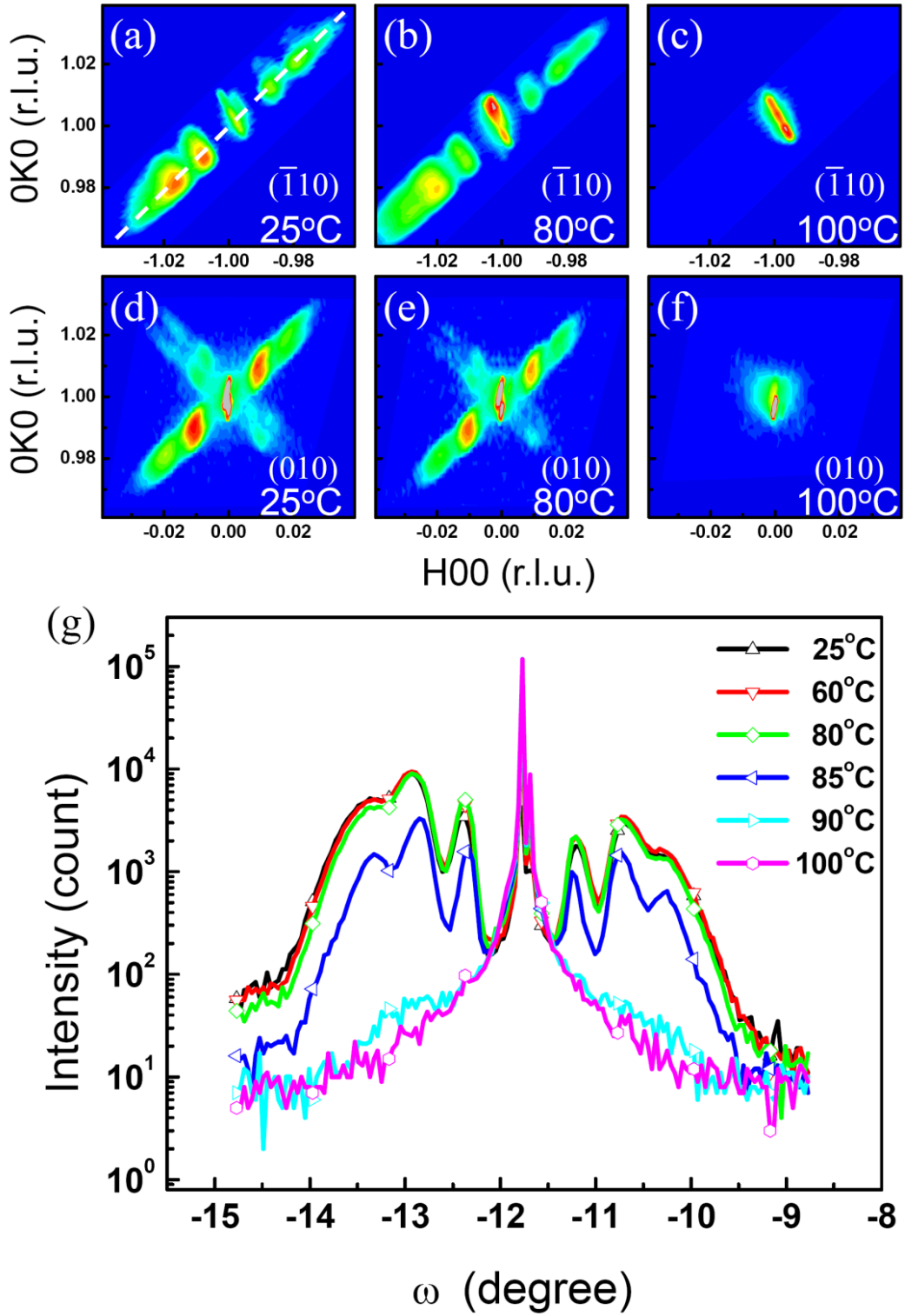

**Figure 5.** (color online) (a-c) ($\bar{1}10$) RSMs, (d-f) (010) RSMs and (g) the corresponding line scans of a 12-nm-thick BFO film at different temperatures. The dotted line in (a) indicates the trail of the line scans. (Intensity from low to high: blue-green-yellow-red-grey).

# Supplementary Material

# Periodic elastic nanodomains in ultrathin tetrogonal-like BiFeO$_3$ films

Zhenlin Luo,[1,*,†] Zuhuang Chen,[2, §,†] Yuanjun Yang,[1] Heng-Jui Liu,[3] Chuanwei Huang,[2] Haoliang Huang,[1] Haibo Wang,[1] Meng-Meng Yang,[1] Chuansheng Hu,[1] Guoqiang Pan,[4] Wen Wen,[5] Xiaolong Li,[5] Qing He,[5] Thirumany Sritharan,[2] Ying-Hao Chu,[3] Lang Chen,[2, 6‡] and Chen Gao[1,#]

[1]*National Synchrotron Radiation Laboratory and CAS Key laboratory of Materials for Energy Conversion, Department of Materials Science and Engineering, University of Science and Technology of China, Hefei 230026, China*

[2]*School of Materials Science and Engineering, Nanyang Technological University, Singapore 639798, Singapore*

[3]*Department of Materials Science and Engineering, National Chiao Tung University, Hsinchu 30010, Taiwan*

[4]*National Synchrotron Radiation Laboratory, University of Science and Technology of China, Hefei 230026, China*

[5]*Shanghai Synchrotron Radiation Facility, Shanghai Institute of Applied Physics, Chinese Academy of Sciences, Shanghai 201204, China*

[6]*Department of Physics, South University of Science and Technology of China, Shenzhen, China, 518055, P.R. China*

† Z. L. Luo and Z. H. Chen contributed equally to this work.

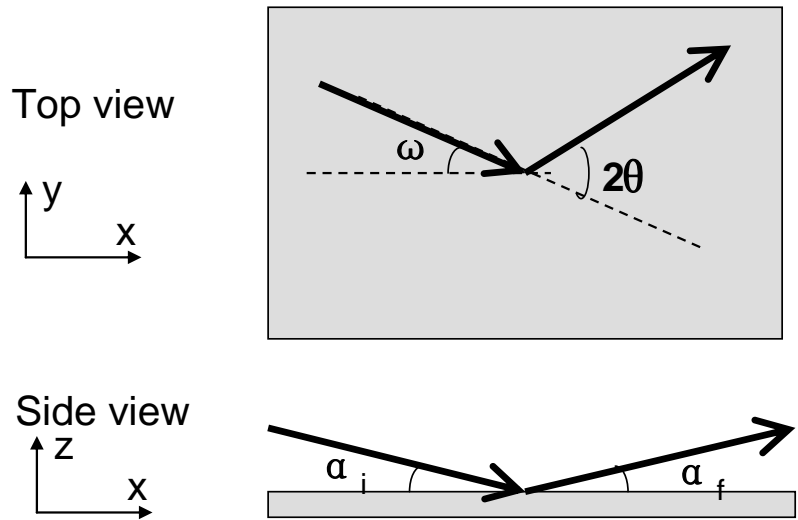

**Fig.S1 Schematic of Grazing Incidence X-ray Diffraction (GIXRD)**

When conducting GIXRD, the incident angle $α_i$ and exit angle $α_f$ are often kept at the same value that is below the critical angle of the surface materials. So, the external reflectivity is almost 100%, and only the evanescent waves penetrate into the films. These evanescent waves have a penetration depth of several nano-meters and thus are suitable for characterizing ultrathin film's in-plane structure by using in-plane ω-2θ geometry, and thereby avoiding the substrate signals. In our GIXRD studies, the incident and exit angles were fixed at nominal 0.2°, thus the attenuation depth in BFO should be ~1.1nm for x-ray at 8 kV. However, due to twins in the LAO substrate, the real incident angle on the whole film could not be exactly 0.2°, which may lead to LAO substrate reflections being picked up in the GIXRD patterns.

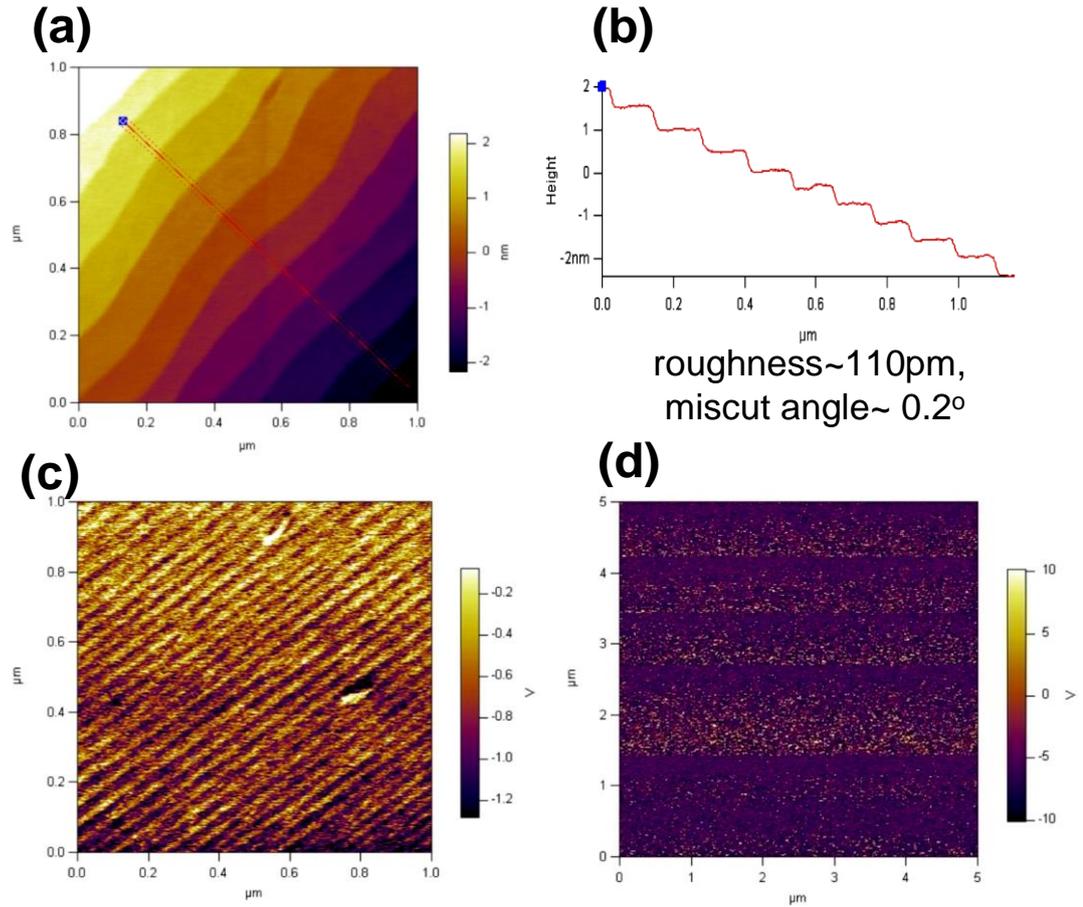

Fig.S2 (a) AFM topography, (c) in-plane and (d) out-of-plane PFM image of the 24-nm-thick film, (b) Line scan profile along the line drawn in (a).

The atomic force microscopy (AFM) and piezoelectric force microscopy (PFM) investigations were performed on an Asylum Research MFP-3D atomic force microscope using TiPt-coated Si tips (DPE18, MikroMasch). The PFM images have been recorded with the tip cantilever pointing along <100> direction.[1]

The high-quality film is demonstrated by AFM topography (Fig. S2a), which reveals that the surface of the film has single-unit-cell step structure (Fig. S2b). In-plane PFM image (Fig.S2(c)) reveals the appearance of periodic stripe domains with domain walls along [110]. The domain periodicity is ~45nm by PFM, which is consistent with the average value of 50nm obtained by GIXRD. Out-of-plane PFM images (Fig.S2(d)) shows uniform contrast, indicating the polarization in the $M_C$ nanodomains have the same out-of-plane component.

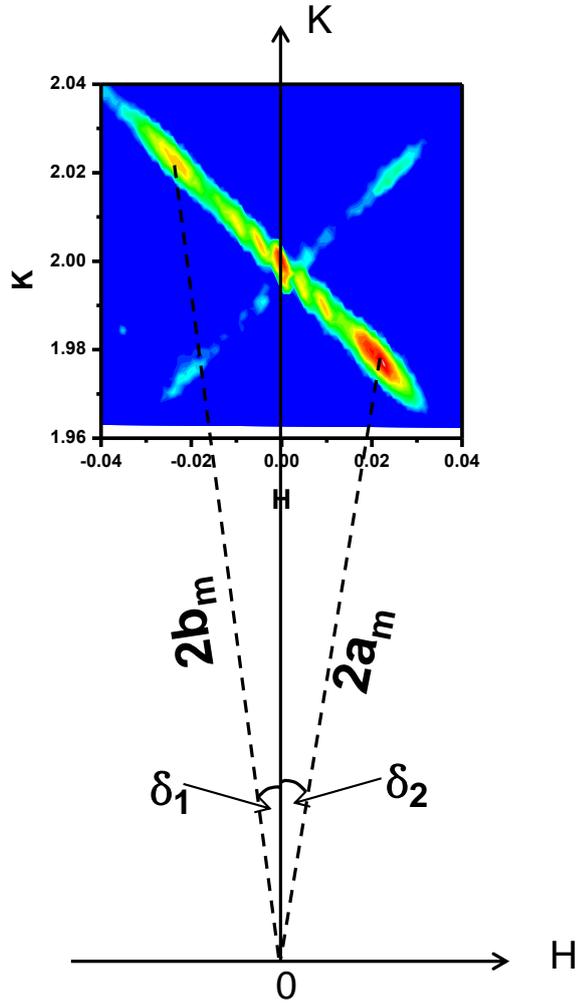

Fig. 3. Illustration of calculation of in-plane rotation angle and lattice constants of $M_C$ BFO phase through (020) RSM of 24-nm-thick BFO film. $a_m$, $b_m$ are the lattice constants of reciprocal lattice unit (r.l.u.) of BFO and $\delta$ is the in-plane rotation angle. (Intensity from low to high: blue-green-yellow-red-grey)

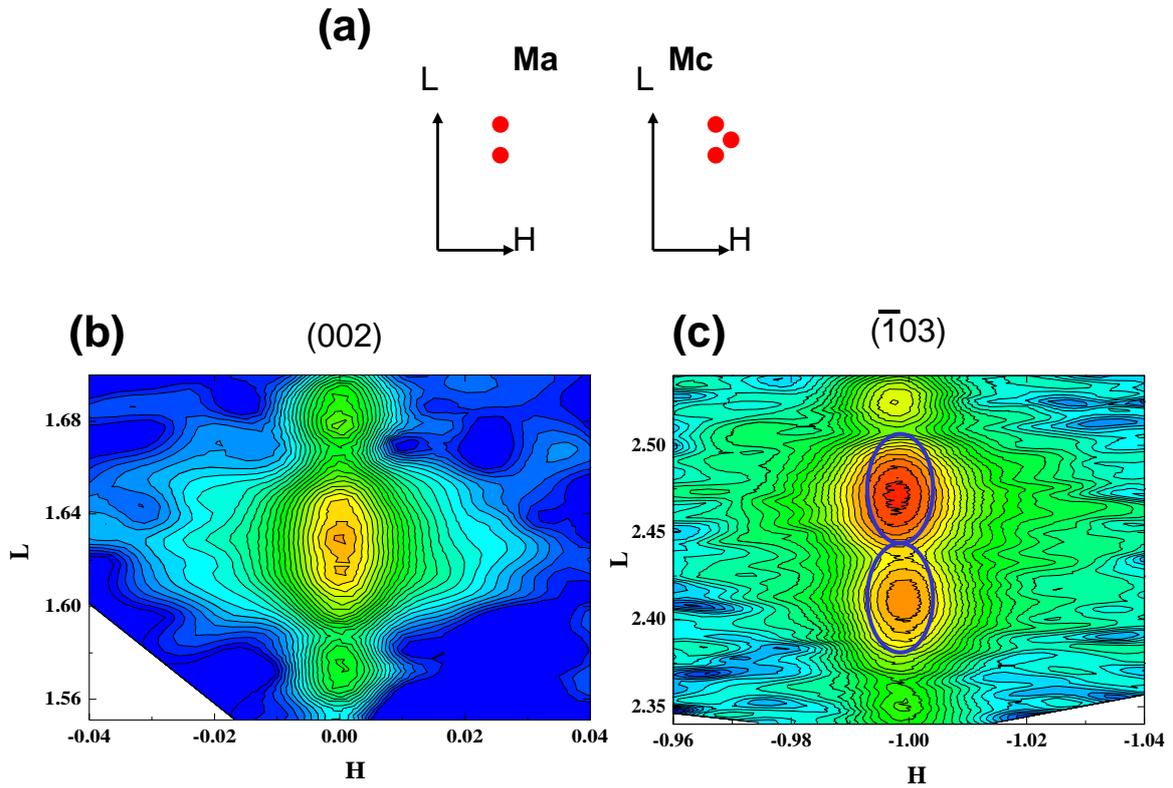

Fig.S4 (a) Sketches of (h0l) (h≠0) RSMs of monoclinic $M_A$ and monoclinic $M_C$ phases, respectively. (b) (002) RSM and (c) $(\bar{1}03)$ RSM of 12-nm-thick BFO film at 100°C. (Intensity from low to high: blue-green-yellow-red).

Profile of the $(\bar{1}03)$ RSM of the 12-nm-thick BFO film indicates that the film is T-like monoclinic $M_A$ phase at 100°C, consistent with previous reports. [2-4]

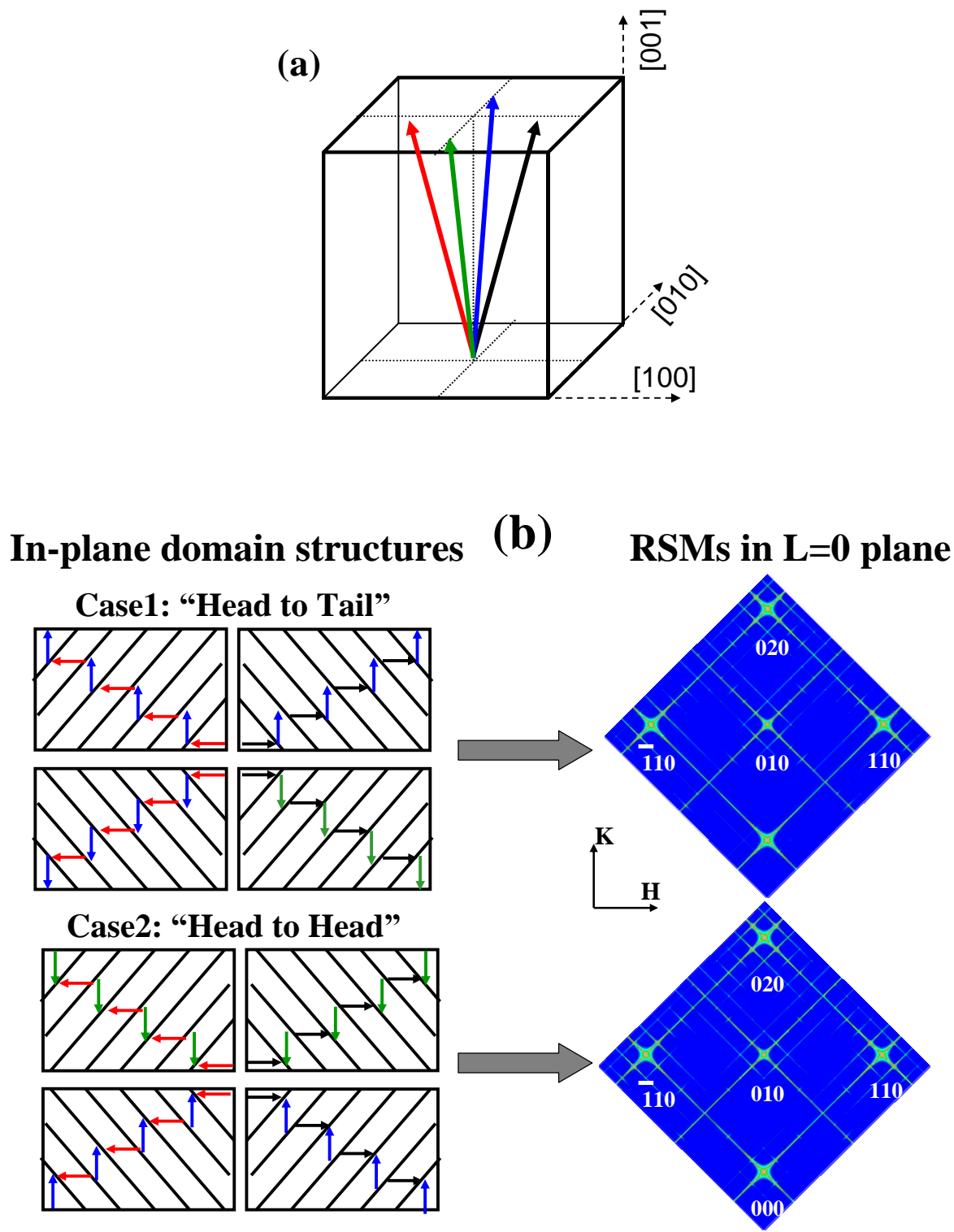

Fig. S5, (a) Sketch of the four domain variants in the monoclinic $M_C$ phase. Arrows represent polarization vectors. (b) Illustration of two possible cases of periodic domain structures with domain walls along <110> directions for $M_C$ film in real space (left) and the corresponding simulated RSMs in L=0 plane (right). As seen, each case assumes the presence

of four different domain structures due to the four-fold in-plane symmetry of (001)-oriented pseudo-cubic substrate. For the simulation it is assumed that the four domain structures are equal in population. "Head" and "Tail" refer to the two different end points of the polarization vector.

In a ferroelectric material, there is a difference in unit cell structure factor between unit cells with different polarization directions. If the domains with different polarization vectors order periodically, as shown in Fig.S5 (b), the resulting difference in structure factor between adjacent domains leads to extra peaks (satellites). In the direction perpendicular to the domain walls, a pair of adjacent domains can be considered as a unit cell of the superstructure. The structure factor of a unit cell is given by:

$$F(q) = \sum_{j=1}^{N} f_j \cdot e^{i(x_j \cdot q_x + y_j \cdot q_y + z_j \cdot q_z)}$$

where $f_j$ are the atomic scattering factors and the sum is over the atoms of one unit cell.

For the simulation conducted here, a supercell consisting of 5 by 5 basic rectangle units of $BiFeO_3$ is adopted and the in-plane shift component of ion atom from the center of unit cell is assumed to be 0.15 Å along in-plane polarization component.

The simulation results show that: 1) for the "head to head" case, satellite along both $[\bar{1}\bar{1}0]$ and $[110]$ directions should appear in the $(\bar{1}10)$ RSM; 2) for the "head to tail" case, only satellite along $[110]$ direction will appear in $(\bar{1}10)$ RSM. Our experimental results are in good agreement with the head to tail domain structure.

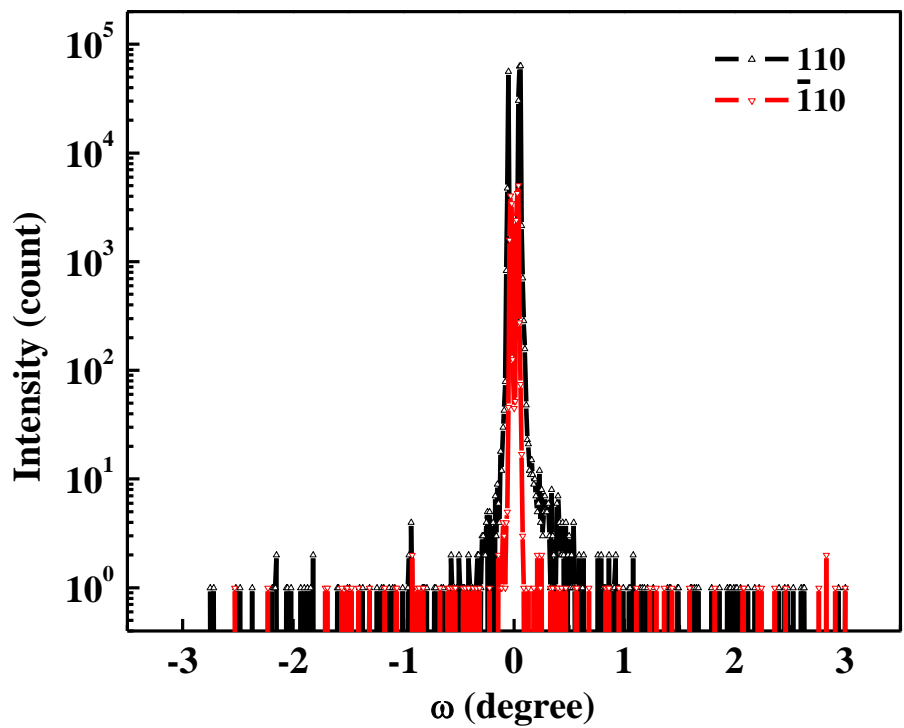

Fig.S6 Rocking curves of (110) and ($\bar{1}10$) reflections of bare LAO substrate after similar heat treatment (2θ=33.4°). No satellite peaks due to periodic modulation were observed. GIXRD geometry was adopted here and the reflection peaks in the cures were realigned at ω=0 for display.)